\documentclass[reprint,amsmath,amssymb,aip]{revtex4-2}

\usepackage{amsmath,amssymb,amsfonts}
\usepackage{url}
\usepackage{tcolorbox}
\usepackage{graphicx}
\usepackage{hyperref}
\usepackage{bm}
\hypersetup{hidelinks}
\DeclareMathOperator{\vecop}{vec}
\newcommand{\fig}[3]{\includegraphics[width=#2]{#1}\hspace{0.5em}#3}
\newcommand{\figline}[1]{\vskip6pt\noindent\hbox to\hsize{#1}\vskip6pt}

\allowdisplaybreaks[4]

\newcommand{\bc}[1]{\mbox{\boldmath $\mathcal{#1}$}}

\begin{document}

\title[Tensor Evolution for Source Depth]{Dynamic Mode Decomposition by Tensor Evolution for Source Depth}
\author{Wenqian Wu}
\affiliation{Polytechnic Institute, Zhejiang University, Hangzhou, 310015, China}

\author{Xiaohong Yang}
\affiliation{Science and Technology on Sonar Laboratory, Hangzhou Applied Acoustics Research Institute, Hangzhou, 310023, China}

\author{Guangying Zheng}
\affiliation{Science and Technology on Sonar Laboratory, Hangzhou Applied Acoustics Research Institute, Hangzhou, 310023, China}

\author{Lei Cheng}
\affiliation{College of Information Science and Electronic Engineering, Zhejiang University, Hangzhou, 310027, China}

\author{Peter Gerstoft}
\affiliation{Acoustic Technology, Department of Electrical and Photonics Engineering, Technical University of Denmark, 2800 Kgs. Lyngby, Denmark}

\date{}

\begin{abstract}
	
Depth estimation of shallow acoustic sources using a deep-ocean near-bottom vertical line array relies on depth-sensitive interference, which appears as an oscillatory structure in the frequency-angle domain in broadband matched beam-intensity processing (MBIP). For low-SNR, the nonlinear feature extraction in MBIP transforms array-domain noise into feature-domain distortions, making depth estimation fragile. Addressing this issue, we propose a tensor evolution-based depth estimation (TEDS) method inspired by dynamic mode decomposition (DMD), which interprets a sequence of broadband MBIP surfaces as the output of a dynamic system from a time-varying autoregressive operator. The operator is represented by a low-rank Tucker decomposition, which separates coherent feature modes and their temporal modes from incoherent noise. The target depth is inferred via Fourier summation applied to a dominant feature mode. Deep-ocean numerical experiments demonstrate that TEDS improves robustness at low-SNR, consistently outperforming matched field processing and broadband MBIP.

\end{abstract}

\maketitle

\begin{center}
\footnotesize
Copyright 2026 Acoustical Society of America. This article may be downloaded for personal use only. Any other use requires prior permission of the author and the Acoustical Society of America.

The following article appeared in W. Wu, X. Yang, G. Zheng, L. Cheng, and P. Gerstoft, ``Dynamic mode decomposition by tensor evolution for source depth,'' \textit{J. Acoust. Soc. Am.} \textbf{160}(3), 2356--2369 (2026), and may be found at \url{https://doi.org/10.1121/10.0046532}.
\end{center}

\section{\label{sec:1} Introduction}

Localizing sources remains a fundamental problem in ocean acoustics.\cite{baggeroer1988matched} Among the parameters involved in passive localization, source depth is particularly important yet difficult to estimate accurately.\cite{yang1990modal} Matched field processing (MFP) is the standard approach, comparing measured acoustic fields with model-predicted replicas.\cite{baggeroer1988matched} Its performance depends on imperfectly known environmental parameters, such as sound-speed profile and seabed properties. Modest mismatch causes substantial degradation.\cite{tolstoy1989sensitivity,hunter2021range} The computational burden of replica generation also limits its practicality in real-time applications.\cite{michalopoulou2021matched,jenkins2023bayesian}

This vulnerability becomes more pronounced for low-SNR, since the ambiguity surface is distorted and its dominant peak shifts away from the true source location.\cite{tolstoy1989sensitivity,hunter2021range} Many approaches based on replica matching or processed-feature matching infer source depth from isolated observations.\cite{baggeroer1988matched,yang1990modal,zheng2020matched,zhou2022target} Under low SNR, once the discriminative structure in a single snapshot becomes weak, the estimate is fragile. What is lacking is a representation that accumulates weak but consistent depth-related information over time while suppressing incoherent fluctuations.

This motivates a dynamic modeling viewpoint. For a fixed source depth under stable propagation geometry, the depth-related structure evolves coherently over time, whereas noise-induced distortions fluctuate irregularly. A representation that explicitly models this evolution is therefore better suited to separating physically meaningful structure from incoherent noise than one based solely on instantaneous feature appearance.

Dynamic mode decomposition (DMD) and related Koopman-based methods provide a natural point of reference for this perspective.\cite{schmid2010dynamic,tu2013dynamic,williams2015data,kutz2016dynamic,brunton2022datascience} They represent high-dimensional observations by a small number of dominant feature modes and their temporal evolution. In the present problem, however, the feature sequence is nonstationary: source motion, time-varying noise, and nonlinear processing make a single time-invariant evolution operator restrictive. This calls for a time-varying extension that preserves coherent feature evolution while accommodating moderate nonstationarity.

Motivated by this observation, we propose a tensor evolution-based depth estimation (TEDS) method inspired by DMD. TEDS interprets a sequence of observations as the output of a dynamic system governed by a time-varying autoregressive operator. This operator is organized as a tensor and represented by a low-rank Tucker decomposition, yielding a compact set of feature modes and corresponding temporal modes. In this way, coherent depth-related structure is concentrated into a few dominant modes, while incoherent noise-driven variability is suppressed. Source depth is then inferred from a dominant feature mode.

The proposed tensor structure regularizes the evolution process rather than aggregating individual observations. This formulation differs from most tensor-based studies in ocean acoustics, which focus on static tasks such as sound-speed-field representation, inversion, and reconstruction and exploit low-rank structure in the data itself.\cite{cheng2022tensor,chen2022tensor,li2023striking,li2024zero}

In this work, the dynamic viewpoint is applied to broadband matched beam-intensity processing (MBIP) features obtained using a deep-ocean near-bottom vertical line array (VLA).\cite{zheng2020matched,zhou2022target} When a shallow source is observed on a VLA, the received field is dominated by a direct reliable acoustic path (RAP) arrival and a surface-reflected arrival.\cite{mccargar2013depth,kniffin2016performance} Their interference produces a depth-dependent oscillatory structure that is governed by propagation geometry and is therefore less sensitive to environmental uncertainty than full-field replica matching.\cite{baggeroer1988matched,mccargar2013depth,kniffin2016performance} This deep-ocean geometry also benefits from reduced propagation loss and weaker surface-noise contamination, favoring the extraction of depth-sensitive interference.\cite{wenz1962acoustic,gaul2007ambient} Under low SNR, however, this oscillatory structure is still obscured, and nonlinear processing such as beamforming and subsequent feature matching further blurs the underlying periodicity, leading to unstable and biased depth estimation.

The remainder is organized as follows. Section~\ref{sec:2} presents the background on depth estimation by first reviewing RAP geometry and MBIP observables and then introducing the dynamic modeling viewpoint. Section~\ref{sec:3} develops the proposed TEDS approach, including tensor evolution-based autoregression for MBIP features and the associated algorithm. Numerical results and discussion are given in Section~\ref{sec:4}. Finally, Section~\ref{sec:5} concludes the paper.

{\bf Notations:}
Lowercase and uppercase bold letters (e.g., $\mathbf{x}$ and $\mathbf{X}$) denote vectors and matrices. Higher-order tensors (order three or higher) are denoted by bold calligraphic letters (e.g., $\bc{A}$). The Kronecker product is denoted by $\otimes$, and the Moore--Penrose pseudoinverse by $(\cdot)^{\dagger}$. The transpose and trace are denoted by $(\cdot)^{T}$ and $\textrm{tr}(\cdot)$. The identity matrix of size $R$ is denoted by $\mathbf{I}_{R}$. For a tensor $\bc{X}$, $\bc{X}_{(n)}$ denotes its mode-$n$ unfolding, and $\times_n$ denotes the mode-$n$ tensor--matrix product. The operator $\mathrm{TV}(\cdot)$ denotes the total variation of a vector.

\section{\label{sec:2} Depth estimation background}

Depth estimation with a deep-ocean near-bottom VLA is commonly approached either through MFP or through RAP-based feature extraction.\cite{baggeroer1988matched,mccargar2013depth,duan2014moving,kniffin2016performance,duan2019performance,zheng2020matched,zhou2022target} The former relies on full-field replica matching, whereas the latter exploits depth-dependent interference in processed observables. The proposed dynamic modeling framework is applied to broadband MBIP features generated under RAP propagation. This section therefore first reviews the RAP geometry and MBIP observables, and then formalizes the dynamic modeling viewpoint underlying the proposed method.

\subsection{\label{subsec:II-A} RAP geometry and MBIP observables}

Previous studies demonstrate RAP for deep-ocean source depth estimation.\cite{mccargar2013depth,duan2014moving,kniffin2016performance,duan2019performance,zheng2020matched,zhou2022target} Consider a VLA with element depths $\{z_j\}_{j=1}^{J}$ and spacing $d$. When the array is deployed below the conjugate depth, the received field is dominated by the direct RAP arrival and a surface-reflected arrival. Because this dual-arrival structure is governed mainly by propagation geometry, it is less sensitive to environmental mismatch than full-field replica matching.\cite{mccargar2013depth,kniffin2016performance}

\begin{figure}[!t]
	\center
	\includegraphics[width=1\columnwidth]{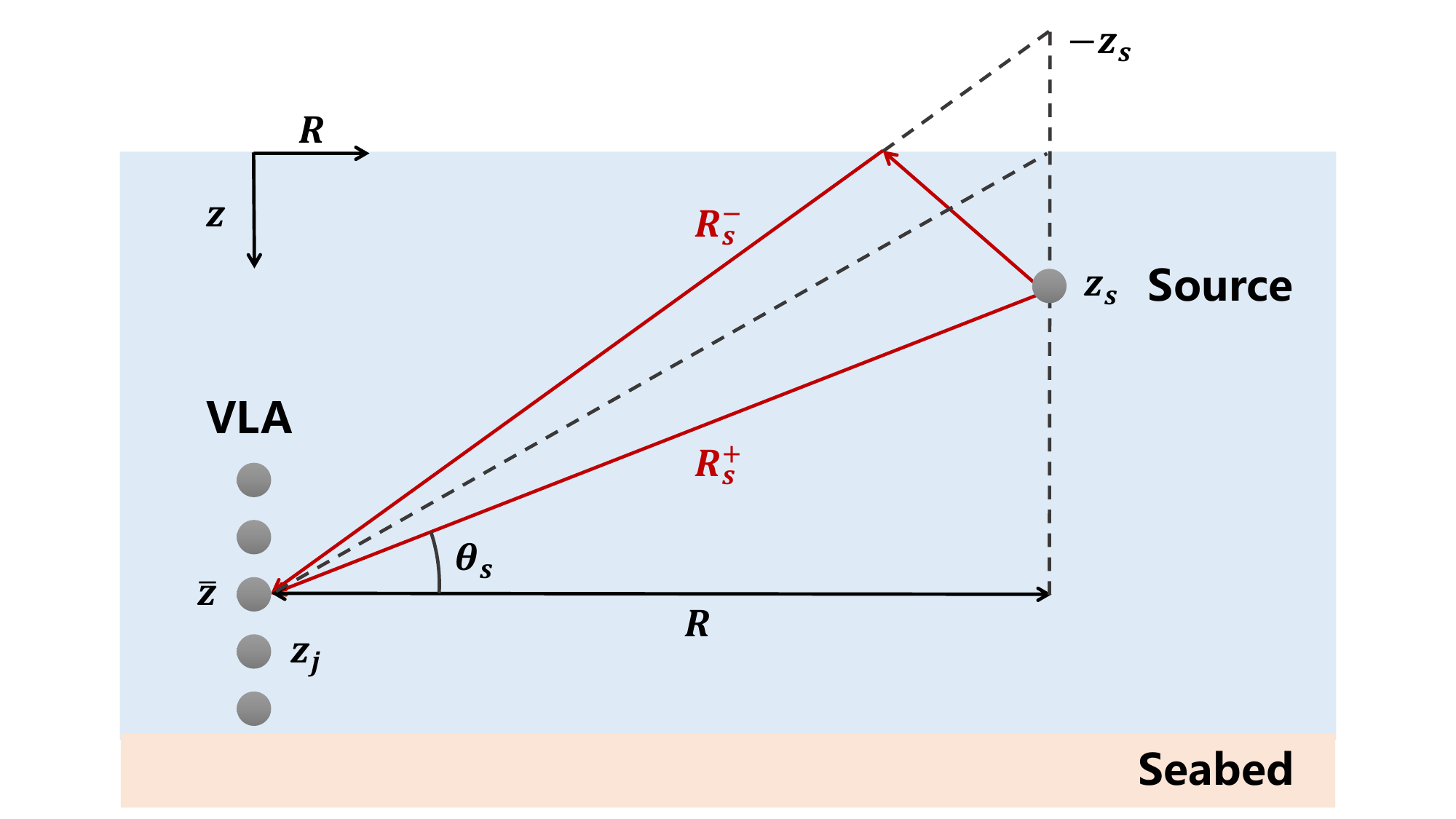}
	\caption{Geometry diagram of sound source and VLA.}
	\label{fig:rap}
	\hrule
\end{figure}

As shown in Fig.~\ref{fig:rap}, for a source at depth $z_s$ and range $R$, the frequency-domain pressure field measured by a hydrophone at depth $z_j$ is\cite{mccargar2013depth,zheng2020matched}
\begin{align}
	p(\omega,z_s,z_j)=S(\omega)\left[\frac{e^{ikR_s^+}}{R_s^+}-\frac{e^{ikR_s^-}}{R_s^-}\right],
	\label{eq:rap_pressure}
\end{align}
where $S(\omega)$ is the source spectrum, $k=k(\omega)$ is the wavenumber, and the lengths of the direct and surface-reflected paths are $R_s^+=\sqrt{R^2+(z_j-z_s)^2}$ and $R_s^-=\sqrt{R^2+(z_j+z_s)^2}$.

Under the deep-ocean geometry, the source grazing angle at the array, whose geometric center is located at depth $\bar{z}$, is approximated by
\begin{align}
	\sin\theta_s \approx \frac{\bar{z}}{\sqrt{R^2+\bar{z}^2}}.
	\label{eq:sin_theta}
\end{align}

Conventional beamforming gives the beam-intensity surface
\begin{align}
	B(\omega,\sin\theta,z_s)=\left|\sum_{j=1}^{J} e^{ik(jd-\bar{z})\sin\theta}\, p(\omega,z_s,z_j)\right|^2,
	\label{eq:beamforming}
\end{align}
whose peak occurs near the steering angle corresponding to the source location. Substituting $\sin\theta=\sin\theta_s$ yields
\begin{align}
	B(\omega,\sin\theta_s,z_s)
	=2\frac{|S(\omega)|^2}{\bar{z}^{2}}\sin^{2}\theta_s
	\left[1-\cos\!\left(2k z_s \sin\theta_s\right)\right].
	\label{eq:interference}
\end{align}
Eq.~\eqref{eq:interference} shows that RAP-induced interference produces a depth-dependent oscillatory structure at the target bearing. The oscillation frequency is determined by the phase term $2k z_s\sin\theta_s$ and therefore carries source-depth information.\cite{mccargar2013depth,zheng2020matched}

MBIP exploits this structure by comparing observed beam-intensity oscillations with those predicted under hypothesized source depths.\cite{zheng2020matched,zhou2022target} In broadband form, the information is aggregated across frequency through the matching coefficient
\begin{equation}
	\begin{aligned}
		M(z)=
		\frac{\int_{\omega_1}^{\omega_2}
			B_{\mathrm{r}}(\omega,\sin\theta,z)\,
			B_{\mathrm{d}}(\omega,\sin\theta_s,z_s)\, d\omega}
		{\sqrt{\int_{\omega_1}^{\omega_2} B_{\mathrm{r}}^2(\omega,\sin\theta,z)\, d\omega}\,
			\sqrt{\int_{\omega_1}^{\omega_2} B_{\mathrm{d}}^2(\omega,\sin\theta_s,z_s)\, d\omega}},
	\end{aligned}
	\label{eq:mbip}
\end{equation}
where $B_{\mathrm{r}}$ and $B_{\mathrm{d}}$ denote the theoretical and observed beam-intensity oscillations. The source depth is estimated by maximizing $M(z)$ over the candidate depth grid.\cite{zheng2020matched,zhou2022target}

The broadband MBIP provides a sequence of beam-intensity surfaces as observables over time. For one source depth, the dominant interference structure remains coherent across successive observations, whereas noise-induced distortions vary irregularly. This distinction motivates a dynamic treatment of MBIP observables rather than a purely snapshot-based one.

\subsection{\label{subsec:II-B} Dynamic modeling viewpoint for robust depth estimation}

MFP formulates passive localization as an inverse problem.\cite{baggeroer1988matched,yang1990modal} For a $J$-element VLA recording the complex pressure $\mathbf{p}\in\mathbb{C}^{J}$ at one frequency, let $\mathbf{u}=[r_s,z_s]^T$ denote the source location, where $r_s$ and $z_s$ are the horizontal range and depth. The field predicted by the forward model $\mathrm{G}$ for a hypothesized source location $\mathbf{u}$ is
\begin{align}
	\mathbf{p}_r=\mathrm{G}(\mathbf{u}).
	\label{eq:obs}
\end{align}

Depth estimation is then performed by comparing the observation $\mathbf{p}_d$ with the replica over a discretized search space $U$. Let $f(\mathbf{u})$ denote the ambiguity surface, taking the conventional Bartlett processing form $f(\mathbf{u}) = \frac{|\mathbf{p}_d^\mathrm{H} \mathbf{p}_r(\mathbf{u})|^2}{\|\mathbf{p}_d\|^2 \|\mathbf{p}_r(\mathbf{u})\|^2}$.\cite{baggeroer2002overview} The estimated source location is obtained by
\begin{align}
	\hat{\mathbf{u}}=\arg\max_{\mathbf{u}\in U} f(\mathbf{u}).
	\label{eq:mfp_est}
\end{align}

MFP treats source range and depth as jointly unknown and does not exploit the RAP geometry, so localization requires a two-dimensional search. Broadband processing improves robustness by combining information across frequency, but it requires more replica generation.\cite{michalopoulou2021matched,jenkins2023bayesian} MFP is inherently sensitive to noise, environmental mismatch, and array uncertainty.\cite{tolstoy1989sensitivity,hunter2021range}

RAP-based methods avoid full-field replica matching, but they still infer depth from a single observation.\cite{mccargar2013depth,duan2014moving,kniffin2016performance,duan2019performance,zheng2020matched,zhou2022target} With reduced interference contrast, both periodicity extraction and feature matching become compromised. 

To formalize the dynamic modeling idea, Let $\mathbf{y}_t\in\mathbb{R}^{N}$ denote the beam-intensity surface in Eq.~\eqref{eq:beamforming} at time $t$, where $N$ is the product of the frequency and angle dimensions. Rather than inferring depth solely from $\mathbf{y}_t$ as in Eqs.~\eqref{eq:interference}--\eqref{eq:mbip}, we regard the sequence $\{\mathbf{y}_t\}_{t=1}^{T}$ as the output of a dynamic system
\begin{align}
	\mathbf{y}_{t+1}=\mathcal{F}_t(\mathbf{y}_t)+\boldsymbol{\epsilon}_t,
	\label{eq:dyn_general}
\end{align}
where $\mathcal{F}_t(\cdot)$ denotes the evolution rule and $\boldsymbol{\epsilon}_t\in\mathbb{R}^{N}$ is a residual term. For a source moving at a constant depth, adjacent MBIP features evolve coherently in time due to the stable RAP geometry, whereas noise-induced distortions vary irregularly. A dynamic representation is therefore better suited to separating coherent depth-related structure from incoherent noise.

DMD provides a reference point.\cite{schmid2010dynamic,tu2013dynamic,williams2015data,kutz2016dynamic,brunton2022datascience} In its standard form, DMD approximates the evolution of successive observations in Eq.~\eqref{eq:dyn_general} by a linear operator
\begin{align}
	\mathbf{y}_{t+1}\approx \mathbf{A}\mathbf{y}_t,
	\label{eq:dmd_basic}
\end{align}
where $\mathbf{A}\in\mathbb{R}^{N\times N}$ is time-invariant. Given the collected observations
\begin{align}
	\mathbf{Y}_1=[\mathbf{y}_1,\ldots,\mathbf{y}_{T-1}], \qquad
	\mathbf{Y}_2=[\mathbf{y}_2,\ldots,\mathbf{y}_{T}],
	\label{eq:dmd_snapshots}
\end{align}
the least-squares estimate of the evolution operator is\cite{schmid2010dynamic,tu2013dynamic}
\begin{align}
	\mathbf{A}_{\mathrm{DMD}}=\mathbf{Y}_2\mathbf{Y}_1^{\dagger}.
	\label{eq:dmd_operator}
\end{align}
DMD modes are obtained from the eigenvectors of $\mathbf{A}_{\mathrm{DMD}}$, or equivalently from those of its low-dimensional projected approximation used in computation. These modes capture the dominant feature modes of the underlying dynamical system.

For the MBIP features, a single time-invariant operator is restrictive. Source motion, time-varying SNR, and feature-domain distortions introduce nonstationarity. We therefore adopt the time-varying extension
\begin{align}
	\mathbf{y}_{t+1}\approx\mathbf{A}_t\mathbf{y}_t,
	\label{eq:dmd_timevarying}
\end{align}
where $\mathbf{A}_t$ varies with time but remains constrained to a structured low-dimensional family. This preserves the mode-based DMD. The corresponding tensor-structured model is developed in Sec.~\ref{sec:3}.

\section{\label{sec:3} TEDS approach}

This section develops the proposed TEDS method. A time-varying autoregressive operator constrained by a low-rank tensor representation is introduced to describe the evolution of MBIP features. On this basis, a constrained optimization objective is formulated for the core tensor and factor matrices, and the resulting problem is solved by alternating updates.

\subsection{\label{subsec:III-A} Tensor evolution-based autoregression for MBIP features}

In broadband MBIP, the feature extracted at each time instant is a beam-intensity surface in the frequency-angle domain, as shown in Eq.~\eqref{eq:beamforming}.\cite{zheng2020matched,zhou2022target} Here, $\mathbf{y}_t\in\mathbb{R}^{N}$ denotes the same feature variable introduced in Eq.~\eqref{eq:dyn_general}.

\begin{figure*}[!t]
	\baselineskip=12pt
	\figline{
		\fig{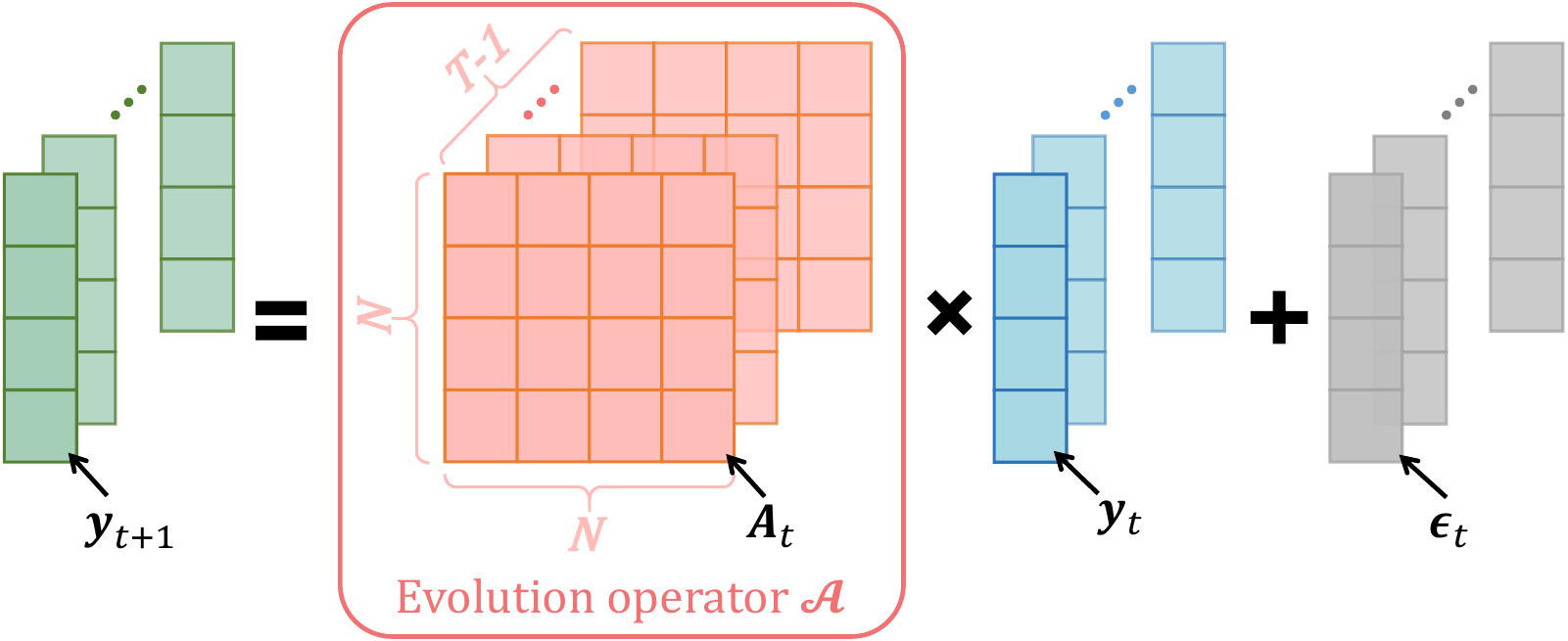}{0.62\textwidth}{(a)} 
		\fig{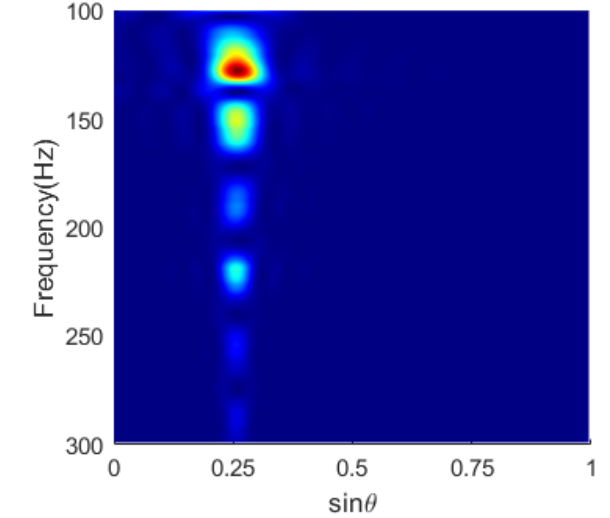}{0.27\textwidth}{(b)} 
	}
	\caption{
		(a) Dynamic evolution of beam-intensity features. The MBIP feature $\mathbf{y}_t$ evolves to $\mathbf{y}_{t+1}$ through a time-varying autoregressive operator, as described by Eq.~\eqref{eq:ar}. In TEDS, this operator is further factorized by a low-rank Tucker representation, yielding the structured evolution model in Eq.~\eqref{eq:ar_tucker}.
		(b) $\mathbf{y}_t$ is the vectorized beam-intensity feature in Eq.~\eqref{eq:beamforming}.}
	\label{fig:dynamic}
	\hrule
\end{figure*}

We model the temporal evolution of the MBIP features by a first-order autoregression
\begin{align}
	\mathbf{y}_{t+1}=\mathbf{A}_t\mathbf{y}_t+\boldsymbol{\epsilon}_t,
	\label{eq:ar}
\end{align}
where $\mathbf{A}_t\in\mathbb{R}^{N\times N}$ is a time-varying evolution operator and $\boldsymbol{\epsilon}_t$ is a residual term, as is illustrated in Fig.~\ref{fig:dynamic}. If $\mathbf{A}_t\equiv \mathbf{A}$ is fixed in time, Eq.~\eqref{eq:ar} reduces to the standard time-invariant DMD.\cite{schmid2010dynamic,tu2013dynamic,kutz2016dynamic}.

The family of operators $\{\mathbf{A}_t\}_{t=1}^{T-1}$ is organized as a third-order tensor $\bc{A}\in\mathbb{R}^{N\times N\times (T-1)}$.\cite{liu2022tensor} To control model complexity and extract coherent feature structure, we impose a low-rank Tucker decomposition with tensor rank $R$, which determines the dimension of the low-rank latent subspace,
\begin{align}
	\bc{A}=\bc{G}\times_1 \mathbf{W} \times_2 \mathbf{V} \times_3 \mathbf{X},
	\label{eq:tucker_tensor}
\end{align}
where $\mathbf{W}\in\mathbb{R}^{N\times R}$ contains the dominant feature modes, $\mathbf{X}\in\mathbb{R}^{(T-1)\times R}$ contains the corresponding temporal modes, $\mathbf{V}\in\mathbb{R}^{N\times R}$ describes the interaction of the current feature with the low-rank operator family, and $\bc{G}\in\mathbb{R}^{R\times R\times R}$ is the core tensor to encode the coupling weights among these three factor matrices and thereby determines how they combine to reconstruct each operator. The same rank $R$ is used so that the factor matrices are all represented in a common $R$-dimensional latent subspace.

The terminology ``modes'' follows DMD and related Koopman-based analysis, in which coherent structure is described by a few dominant modes and their temporal evolution.\cite{schmid2010dynamic,williams2015data,brunton2022datascience} The columns of $\mathbf{W}$ represent dominant spectral-angular modes in the MBIP feature domain, while the columns of $\mathbf{X}$ describe how these modes evolve over time. This yields a DMD-style interpretation while allowing the evolution operator to vary with time.

\begin{figure*}[t]
	\center
	\includegraphics[width=1.75\columnwidth]{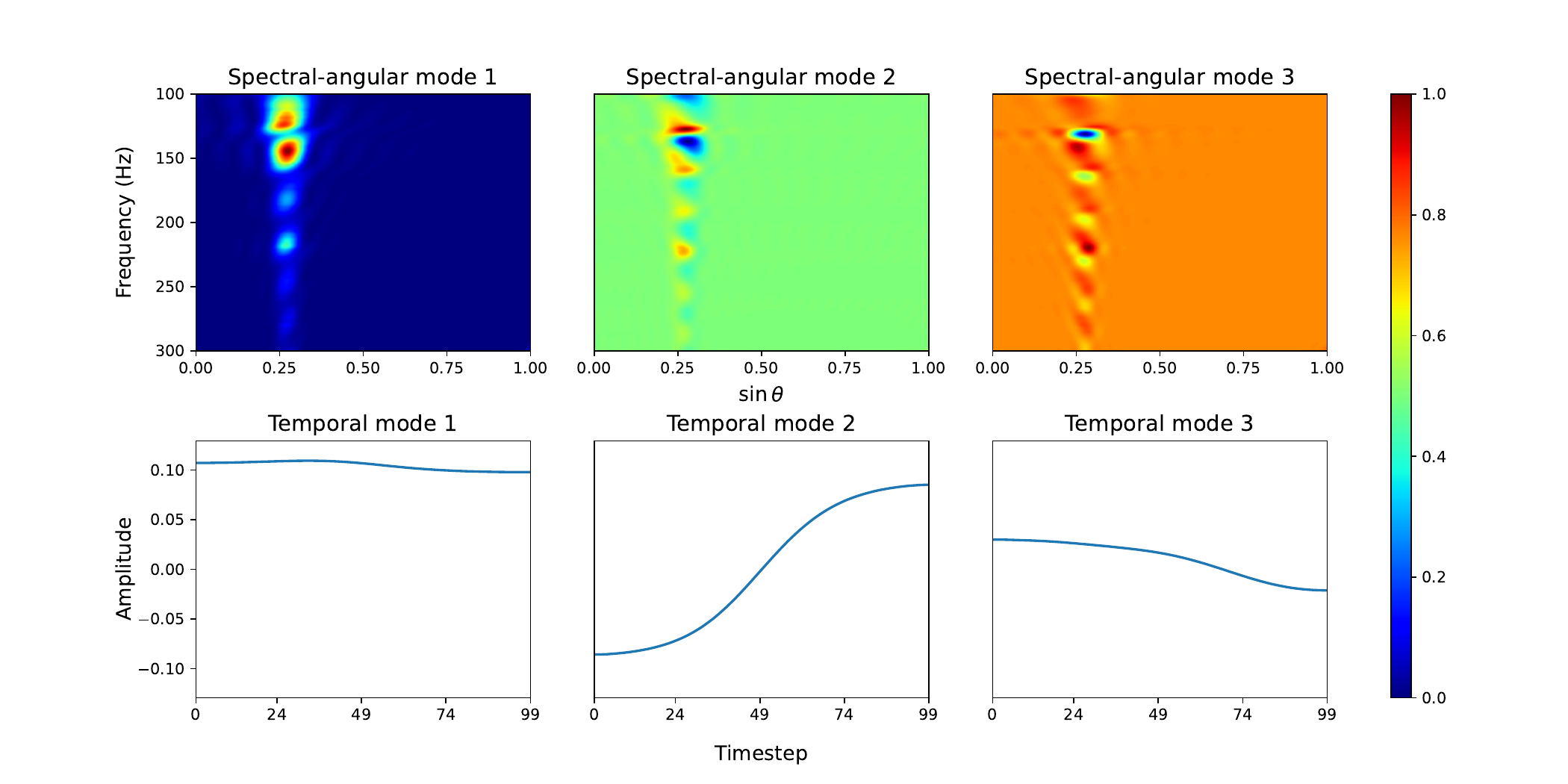}
	\caption{First three spectral-angular modes (top) and corresponding temporal modes (bottom). The spectral-angular modes are the columns of $\mathbf{W}$ and the temporal modes are the columns of $\mathbf{X}$ in Eqs.~\eqref{eq:tucker_tensor}--\eqref{eq:ar_tucker}.}
	\label{fig:plot_modes}
	\hrule
\end{figure*}

To facilitate interpretation, Fig.~\ref{fig:plot_modes} shows the first three extracted spectral-angular modes and temporal modes for a  $100\,\mathrm{m}$-deep source moving uniformly over a $2\,\mathrm{km}$ interval with no nosie. 
Spectral-angular mode~1 captures the mean MBIP information, and its temporal mode is nearly a straight line. 
Spectral-angular mode~2 contains the dominant interference periodicity used for depth estimation. 
Its corresponding temporal mode is approximately sinusoidal, consistent with the interference structure also exhibits periodic variations over time as the source moves.\cite{mccargar2013depth} 
Spectral-angular mode~3 captures a secondary interference component with slightly weaker depth-sensitive information than mode~2. 
The extracted modes are interpreted as a separation of mean component and  depth-sensitive interference components.
Further discussion is in Sec.~\ref{subsec:IV-B}.

Using tensor mode-product identities,\cite{liu2022tensor,kolda2009tensor} the coefficient matrix at time $t$ is
\begin{align}
	\mathbf{A}_t
	=\bc{G}\times_1 \mathbf{W}\times_2 \mathbf{V}\times_3 \mathbf{x}_t^T
	=\mathbf{W}\mathbf{G}\bigl(\mathbf{x}_t\otimes\mathbf{V}^T\bigr),
	\label{eq:tucker}
\end{align}
where $\mathbf{G}\triangleq \bc{G}_{(1)}\in\mathbb{R}^{R\times R^2}$ is the mode-1 unfolding of $\bc{G}$, and $\mathbf{x}_t\in\mathbb{R}^{R}$ is the $t$-th row of $\mathbf{X}$ written as a column vector. Substituting Eq.~\eqref{eq:tucker} into Eq.~\eqref{eq:ar} gives
\begin{align}
	\mathbf{y}_{t+1}
	=\mathbf{W}\mathbf{G}\bigl(\mathbf{x}_t\otimes\mathbf{V}^T\bigr)\mathbf{y}_t+\boldsymbol{\epsilon}_t.
	\label{eq:ar_tucker}
\end{align}

TEDS estimates the factor matrices and the core tensor by solving the constrained optimization problem:
\begin{align}
	\{\mathbf{G},\mathbf{W},\mathbf{V},\mathbf{X}\}
	=
	\arg\min_{\mathbf{G},\mathbf{W},\mathbf{V},\mathbf{X}}\;
	&f(\mathbf{G},\mathbf{W},\mathbf{V},\mathbf{X})
	\notag\\
	\text{s.t.}\quad
	&\mathbf{W}^T\mathbf{W}=\mathbf{I}_R,
	\label{eq:opt_problem}
\end{align}
where
\begin{align}
	f(\mathbf{G},\mathbf{W},\mathbf{V},\mathbf{X})
	=
	&\frac{1}{2}\sum_{t=1}^{T-1}
	\left\|
	\mathbf{y}_{t+1}
	-\mathbf{W}\mathbf{G}\bigl(\mathbf{x}_t\otimes\mathbf{V}^T\bigr)\mathbf{y}_t
	\right\|_2^2 \notag\\
	&+\lambda\sum_{r=1}^{R}\mathrm{TV}(\mathbf{X}_{:,r}),
	\label{eq:opt}
\end{align}
\begin{align}
	\mathrm{TV}(\mathbf{X}_{:,r})
	=\sum_{t=1}^{T-2}\left|\mathbf{X}_{t+1,r}-\mathbf{X}_{t,r}\right|.
	\label{eq:tv}
\end{align}
Here, $\mathrm{TV}(\mathbf{X}_{:,r})$ denotes the total variation of the $r$th temporal mode.\cite{rudin1992nonlinear} In this work, $\lambda$ is selected according to the sensitivity analysis in Appendix~\ref{app:sensitivity}. The orthogonality constraint on $\mathbf{W}$ removes scaling and rotation ambiguities and improves numerical stability and interpretability.\cite{tu2013dynamic,brunton2016discovering,chen2025dynamic} The TV penalty promotes smooth temporal evolution of the dominant modes. In what follows, $\mathbf{G}$, $\mathbf{W}$, $\mathbf{V}$, and $\mathbf{X}$ are updated alternately to solve Eq.~\eqref{eq:opt_problem}.

\subsection{\label{subsec:III-B} Algorithm development}

The optimization is carried out by block coordinate descent:
\begin{equation}
	\left\{
	\begin{aligned}
		\mathbf{G}&=\arg\min_{\mathbf{G}}f(\mathbf{G},\mathbf{W},\mathbf{V},\mathbf{X}),\\
		\mathbf{W}&=\arg\min_{\mathbf{W}^T\mathbf{W}=\mathbf{I}_R}f(\mathbf{G},\mathbf{W},\mathbf{V},\mathbf{X}),\\
		\mathbf{V}&=\arg\min_{\mathbf{V}}f(\mathbf{G},\mathbf{W},\mathbf{V},\mathbf{X}),\\
		\mathbf{X}&=\arg\min_{\mathbf{X}}f(\mathbf{G},\mathbf{W},\mathbf{V},\mathbf{X}).
	\end{aligned}
	\right.
	\label{eq:bcd}
\end{equation}

\noindent $\blacksquare$ {\bf Updating $\mathbf{G}$:}
When $\mathbf{W}$, $\mathbf{V}$, and $\mathbf{X}$ are fixed, define
\begin{align}
	\mathbf{s}_t\triangleq\bigl(\mathbf{x}_t\otimes\mathbf{V}^T\bigr)\mathbf{y}_t\in\mathbb{R}^{R^2}.
	\label{eq:st}
\end{align}

The $\mathbf{G}$-subproblem becomes
\begin{equation}
	\mathbf{G}
	=\arg\min_{\mathbf{G}}
	\frac{1}{2}\sum_{t=1}^{T-1}
	\left\|
	\mathbf{y}_{t+1}-\mathbf{W}\mathbf{G}\mathbf{s}_t
	\right\|_2^2.
	\label{eq:G_sub}
\end{equation}

Since $\mathbf{W}^T\mathbf{W}=\mathbf{I}_R$, left-multiplying the residual by $\mathbf{W}^T$ gives the equivalent problem
\begin{equation}
	\mathbf{G}
	=\arg\min_{\mathbf{G}}
	\frac{1}{2}\sum_{t=1}^{T-1}
	\left\|
	\mathbf{W}^T\mathbf{y}_{t+1}-\mathbf{G}\mathbf{s}_t
	\right\|_2^2.
	\label{eq:G_sub_reduced}
\end{equation}
Eq.~\eqref{eq:G_sub_reduced} is a standard linear least-squares problem. Setting the derivative of the objective with respect to $\mathbf{G}$ to zero yields
\begin{align}
	\sum_{t=1}^{T-1}
	\bigl(\mathbf{G}\mathbf{s}_t-\mathbf{W}^T\mathbf{y}_{t+1}\bigr)\mathbf{s}_t^T
	=\mathbf{0},
\end{align}
which gives
\begin{align}
	\mathbf{G}
	=
	\mathbf{W}^{T}
	\Biggl(\sum_{t=1}^{T-1}\mathbf{y}_{t+1}\mathbf{s}_t^T\Biggr)
	\Biggl(\sum_{t=1}^{T-1}\mathbf{s}_t\mathbf{s}_t^T\Biggr)^{-1},
	\label{eq:updateG}
\end{align}
provided that $\sum_{t=1}^{T-1}\mathbf{s}_t\mathbf{s}_t^T$ is invertible. This update is a linear least-squares fit of the core tensor in unfolded form.

\noindent $\blacksquare$ {\bf Updating $\mathbf{W}$:}
With $\mathbf{s}_t$ in Eq.~\eqref{eq:st}, the $\mathbf{W}$-subproblem is
\begin{align}
	\mathbf{W}
	=\arg\min_{\mathbf{W}^T\mathbf{W}=\mathbf{I}_R}
	\frac{1}{2}\sum_{t=1}^{T-1}
	\left\|
	\mathbf{y}_{t+1}-\mathbf{W}\mathbf{G}\mathbf{s}_t
	\right\|_2^2.
	\label{eq:W_sub}
\end{align}

Expanding the quadratic term yields the equivalent orthogonal Procrustes problem, \cite{schonemann1966generalized}
\begin{align}
	\mathbf{W}
	=\arg\max_{\mathbf{W}^T\mathbf{W}=\mathbf{I}_R}
	\mathrm{tr}\!\left(
	\mathbf{W}^T
	\sum_{t=1}^{T-1}\mathbf{y}_{t+1}\mathbf{s}_t^T\mathbf{G}^T
	\right).
	\label{eq:W_proc}
\end{align}

Let $\mathbf{C}_w\mathbf{\Sigma}_w\mathbf{D}_w^T$ be the singular value decomposition of $\sum_{t=1}^{T-1}\mathbf{y}_{t+1}\mathbf{s}_t^T\mathbf{G}^T$. Then the optimal solution is
\begin{align}
	\mathbf{W}=\mathbf{C}_w\mathbf{D}_w^T.
	\label{eq:updateW}
\end{align}
This update extracts an orthonormal feature subspace that best explains the next-step observations.

\noindent $\blacksquare$ {\bf Updating $\mathbf{V}$:}
When $\mathbf{G}$, $\mathbf{W}$, and $\mathbf{X}$ are fixed, the $\mathbf{V}$-subproblem remains quadratic in $\mathbf{V}$:
\begin{align}
	\mathbf{V}
	=\arg\min_{\mathbf{V}}
	\frac{1}{2}\sum_{t=1}^{T-1}
	\left\|
	\mathbf{y}_{t+1}
	-\mathbf{W}\mathbf{G}\bigl(\mathbf{x}_t\otimes\mathbf{V}^T\bigr)\mathbf{y}_t
	\right\|_2^2 .
	\label{eq:V_sub}
\end{align}

Setting the derivative of Eq.~\eqref{eq:V_sub} with respect to $\mathbf{V}$ to zero yields
\begin{align}
	\sum_{t=1}^{T-1}
	\mathbf{y}_t\mathbf{y}_t^T
	(\mathbf{x}_{t}^T\otimes\mathbf{V})
	\notag
	&\times
	\mathbf{G}^T\mathbf{W}^T\mathbf{W}\mathbf{G}
	(\mathbf{x}_{t}\otimes\mathbf{I}_R)\\
	&=
	\sum_{t=1}^{T-1}
	\mathbf{y}_t\mathbf{y}_{t+1}^T
	\mathbf{W}\mathbf{G}
	(\mathbf{x}_{t}\otimes\mathbf{I}_R),
	\label{eq:vcg}
\end{align}
similar to Eq.~(16) in Ref.~\onlinecite{chen2025dynamic}.

For convenience, let $\bm{\theta}_v:\mathbb{R}^{N\times R}\rightarrow\mathbb{R}^{NR}$ and $\bm{\psi}_v:\mathbb{R}^{N\times R}\rightarrow\mathbb{R}^{NR}$ denote the vectorized left- and right-hand sides of Eq.~\eqref{eq:vcg}. 
Then Eq.~\eqref{eq:vcg} is equivalently written as
\begin{align}
	\bm{\theta}_v(\mathbf{V})=\bm{\psi}_v .
	\label{eq:vcg_vec}
\end{align}
We solve Eq.~\eqref{eq:vcg_vec} using the conjugate gradient method.\cite{shewchuk1994introduction} 
The complete procedure is summarized in \textbf{Algorithm~1}, where $\mathbf{v}_{l+1}$, $\mathbf{r}_{l+1}$, and $\mathbf{d}_{l+1}$ denote the iterate, residual, and search direction, respectively, and $\mathbf{D}_l$ denotes the matrix form of $\mathbf{d}_l$ used as the input to $\bm{\theta}_v(\cdot)$. 
This update adjusts $\mathbf{V}$ to represent how the current observation $\mathbf{y}_t$ participates in the low-rank operator interactions that generate $\mathbf{y}_{t+1}$.

\begin{table}[t]
	\begin{tcolorbox}
		{\textbf{\color{black} Algorithm 1: Conjugate gradient update for $\mathbf{V}$}}
		\\
		\begin{ruledtabular}
			\begin{tabular}{c}
				\leftline{\textbf{Input:} MBIP sequence $\{\mathbf{y}_t\}_{t=1}^{T}$, current $\mathbf{G}$, $\mathbf{W}$, $\mathbf{X}$,} \\
				\leftline{current $\mathbf{V}$, maximum CG iteration $L_V$} \\
				\leftline{\textbf{Initialize:} $l\leftarrow 0$} \\
				\leftline{$\mathbf{v}_0 = \vecop(\mathbf{V})$, $\mathbf{r}_0=\bm{\psi}_v-\bm{\theta}_v(\mathbf{V})$, and $\mathbf{d}_0=\mathbf{r}_0$} \\
				\leftline{Convert $\mathbf{d}_0$ into $\mathbf{D}_0$} \\
				\leftline{\textbf{while} $l<L_V$ \textbf{do}} \\
				\leftline{\ \ \  $\alpha_l=\dfrac{\mathbf{r}_l^T\mathbf{r}_l}{\mathbf{d}_l^T\bm{\theta}_v(\mathbf{D}_l)}$} \\
				\leftline{\ \ \ $\mathbf{v}_{l+1}=\mathbf{v}_l+\alpha_l\mathbf{d}_l$} \\
				\leftline{\ \ \  $\mathbf{r}_{l+1}=\mathbf{r}_l-\alpha_l\bm{\theta}_v(\mathbf{D}_l)$} \\
				\leftline{\ \ \ $\beta_l=\dfrac{\mathbf{r}_{l+1}^T\mathbf{r}_{l+1}}{\mathbf{r}_l^T\mathbf{r}_l}$} \\
				\leftline{\ \ \ $\mathbf{d}_{l+1}=\mathbf{r}_{l+1}+\beta_l\mathbf{d}_l$} \\
				\leftline{\ \ \ Convert $\mathbf{d}_{l+1}$ into $\mathbf{D}_{l+1}$} \\
				\leftline{\ \ \ $l\leftarrow l+1$} \\
				\leftline{Convert $\mathbf{v}_{l}$ into $\mathbf{V}$} \\
				\leftline{\textbf{Return:} $\mathbf{V}$}
			\end{tabular}
		\end{ruledtabular}
	\end{tcolorbox}
	\label{algorithm1}
\end{table}

\noindent $\blacksquare$ {\bf Updating $\mathbf{X}$:}
The matrix $\mathbf{X}\in\mathbb{R}^{(T-1)\times R}$ contains the temporal modes. 
Its $t$-th row $\mathbf{x}_t^T$ determines the operator at time $t$ through Eq.~\eqref{eq:tucker}. 
To promote smooth temporal evolution, we solve
\begin{equation}
	\arg\min_{\mathbf{X}}
	\frac{1}{2}\sum_{t=1}^{T-1}
	\left\|
	\mathbf{y}_{t+1}
	-\mathbf{W}\mathbf{G}\bigl(\mathbf{x}_t\otimes\mathbf{V}^T\bigr)\mathbf{y}_t
	\right\|_2^2
	+\lambda\sum_{r=1}^{R}\mathrm{TV}(\mathbf{X}_{:,r}),
	\label{eq:updateX}
\end{equation}
where
\begin{align}
	\mathrm{TV}(\mathbf{X}_{:,r})
	=
	\sum_{t=1}^{T-2}
	\sqrt{
		\left(\mathbf{X}_{t+1,r}-\mathbf{X}_{t,r}\right)^2
		+\epsilon_{\mathrm{TV}}
	}.
\end{align}
Here, $\epsilon_{\mathrm{TV}}$ is a small smoothing constant that makes the TV penalty differentiable for gradient-based optimization.

Since the TV term couples adjacent rows, the update is carried out on the whole matrix $\mathbf{X}$ rather than row by row.\cite{rudin1992nonlinear} 
Eq.~\eqref{eq:updateX} is minimized in PyTorch using Adam for $L_X$ iterations with learning rate $\eta_X$.\cite{paszke2019pytorch,kingma2015adam}

\subsection{\label{subsec:III-C} Algorithm summary}

The tensor rank $R$ controls the dimension of the latent evolution subspace. 
In this work, $R=3$ is used, as justified in Appendix~\ref{app:sensitivity}. Before the alternating updates, the factor matrices and the core tensor are initialized using a rank-$R$ truncated singular value decomposition of the MBIP snapshot matrix, see Appendix~\ref{app:algorithm_parameters}. The proposed algorithm outputs the spectral-angular modes $\mathbf{W}$, temporal modes $\mathbf{X}$, together with $\bc{G}$ and $\mathbf{V}$.

For depth estimation, one column of $\mathbf{W}$ is selected as the depth-sensitive mode and reshaped into a spectral-angular surface $\mathbf{B}$. Mode~1 represents the mean spectral-angular energy, while mode~2 contains the dominant depth-sensitive interference pattern. When the tracking duration is longer, mode~3 provides a cleaner depth-sensitive component, see Sec.~\ref{subsec:IV-E}. After mode selection, the target-bearing slice $\mathbf{b}$ in Eq.~\eqref{eq:interference} is extracted from the column with the largest integrated modal amplitude. 
The Fourier summation response over candidate depth $z$ is defined as
\begin{equation}
	\mathcal{D}(z)
	=
	\left|
	\sum_{m=1}^{M}
	b(\omega_m)
	e^{-i2k(\omega_m)z\sin\theta^\star}
	\right|,
	\label{eq:fourier_sum}
\end{equation}
and the source depth is estimated by
\begin{equation}
	\hat z
	=
	\arg\max_{\mathbf{z}\in Z}
	\mathcal{D}(z).
	\label{eq:max_F}
\end{equation}
The complete procedure is summarized in \textbf{Algorithm~2}.

\begin{table}[t]
	\begin{tcolorbox}
		{\textbf{\color{black} Algorithm 2: TEDS for target depth estimation}}
		\\
		\begin{ruledtabular}
			\begin{tabular}{c}
				\leftline{\textbf{Input:} MBIP sequence $\{\mathbf{y}_t\}_{t=1}^{T}$} \\
				\leftline{\textbf{Parameters:} $R=3$, $\lambda=0.1$, $L=50$, $L_V=100$,}\\
				\leftline{\ \ \ $L_X=100$,$\eta_X=5\times10^{-3}$, $\epsilon_{\mathrm{TV}}=10^{-6}$} \\
				\leftline{\textbf{Initialize:} $l\leftarrow 0$} \\
				\leftline{\ \ \ Initialize $\mathbf{W}$, $\mathbf{V}$ via Eq.~\eqref{ini_w_v}} \\
				\leftline{\ \ \ Initialize $\mathbf{X}$, Eq.~\eqref{ini_x}} \\
				\leftline{\ \ \ Initialize $\mathbf{G}$, Eq.~\eqref{eq:updateG}} \\
				\leftline{\textbf{while} $l<L$ \textbf{do}} \\
				\leftline{\ \ \  Update $\mathbf{G}$, Eq.~\eqref{eq:updateG}} \\
				\leftline{\ \ \ Update $\mathbf{W}$, Eq.~\eqref{eq:updateW}} \\
				\leftline{\ \ \ Update $\mathbf{V}$ using \textbf{Algorithm~1}} \\
				\leftline{\ \ \ Update $\mathbf{X}$, Eq.~\eqref{eq:updateX} using Adam} \\
				\leftline{\ \ \ $l\leftarrow l+1$} \\
				\leftline{Select the depth-sensitive mode and reshape it into $\mathbf{B}$} \\
				\leftline{Find the target-bearing column by} \\
				\leftline{\qquad $j^\star=\arg\max\limits_j
					\sum_m
					\left|
					\mathbf{B}(\omega_m,\sin\theta_j)
					\right|$} \\
				\leftline{Set $\sin\theta^\star=\sin\theta_{j^\star}$ and extract} \\
				\leftline{\qquad $\mathbf{b}
					=
					[
					\mathbf{B}(\omega_1,\sin\theta^\star),
					\ldots,
					\mathbf{B}(\omega_M,\sin\theta^\star)
					]^T$} \\
				\leftline{Estimate depth via Eq.~\eqref{eq:max_F}}\\
				\leftline{\textbf{Return:} $\hat z$}
			\end{tabular}
		\end{ruledtabular}
	\end{tcolorbox}
	\label{algorithm2}
\end{table}

\section{\label{sec:4} Numerical results and discussion}

\subsection{\label{subsec:IV-A} Simulation settings}

\begin{figure}[!t]
	\baselineskip=12pt
	\figline{
		\fig{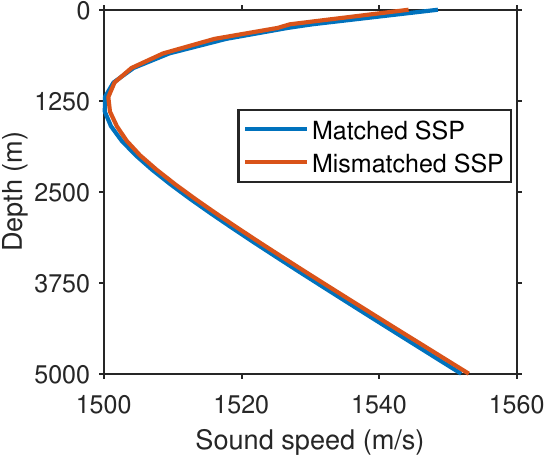}{0.23\textwidth}{(a)} \label{fig:ssp}
		\fig{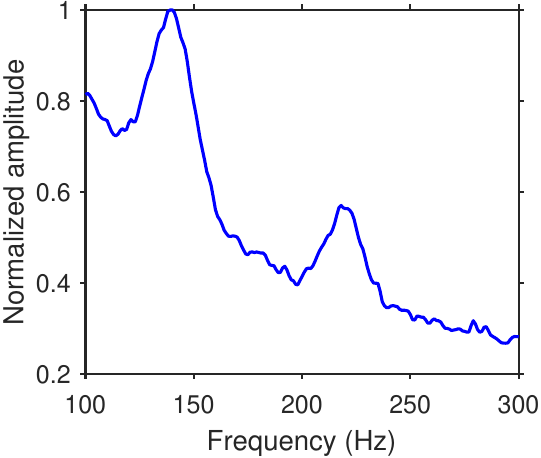}{0.23\textwidth}{(b)} \label{fig:source_spectrum}
	}
	\caption{\label{fig:simulation}
		(a) Matched and mismatched Munk sound-speed profiles.
		(b) Normalized simulation source spectrum.}
	\hrule
\end{figure}

All simulations are carried out for a 32-element VLA with center depth $4900\,\mathrm{m}$ and inter-element spacing $5\,\mathrm{m}$. The nominal environment is Munk profile, see Fig.~\ref{fig:ssp}, with a fluid half-space seabed of sound speed $1600\,\mathrm{m/s}$, density $1.8\,\mathrm{g/cm^3}$, and attenuation $0.8\,\mathrm{dB/(m kHz)}$. Replicas are generated with KRAKEN.\cite{porter1992kraken} The signal band is $100$--$300\,\mathrm{Hz}$, and the source depth varies from $60$ to $200\,\mathrm{m}$ in steps of $20\,\mathrm{m}$.

As indicated by Eq.\eqref{eq:interference}, the periodic interference depends on the source spectrum $S(\omega)$. The synthetic source spectrum has a broadband component with power-law decay to represent low-frequency-dominated energy and two superimposed narrowband components to emulate representative tonal features. \cite{jensen2011computational,wenz1962acoustic} The spectrum was normalized to unit peak amplitude and used as a frequency-dependent weighting, see Fig.~\ref{fig:source_spectrum}. 

Additive noise is introduced at the array-element level. Let $\mathbf{p}(\omega)\in\mathbb{C}^{J}$ denote the noise-free array snapshot and $\mathbf{n}(\omega)\in\mathbb{C}^{J}$ the additive noise. The noisy observation is
\begin{align}
	\mathbf{p}_{\mathrm{noisy}}(\omega)=\mathbf{p}(\omega)+\mathbf{n}(\omega).
	\label{eq:noisy_array}
\end{align}
The signal-to-noise ratio (SNR) is defined as
\begin{align}
	\mathrm{SNR}
	=10\log_{10}
	\frac{\sum_\omega \|\mathbf{p}(\omega)\|_2^2}
	{\sum_\omega\|\mathbf{n}(\omega)\|_2^2}
	\quad \mathrm{dB}.
	\label{eq:snr_def}
\end{align}
The tested conditions are noise-free, $-5$, $-10$, and $-15\,\mathrm{dB}$.

For MFP, both matched and mismatched cases are considered. In the matched case, replicas are generated with the same environment used in the forward simulation. In the mismatched case, replicas are computed using the red SSP in Fig.~\ref{fig:ssp}, together with array tilt and element-dependent gain/phase perturbations, so as to mimic practical environmental and system mismatch. The estimate is obtained by global grid search over the full ambiguity surface defined in Sec.~\ref{sec:2}. In the broadband implementation, 40 uniformly selected frequencies are used in the $100$--$300\,\mathrm{Hz}$ band.

The source is tracked over a $2\,\mathrm{km}$ interval starting from range $16\,\mathrm{km}$, with 200 broadband MBIP observations. Additional results for tracking durations from $1$ to $5\,\mathrm{km}$ are given in Sec.~\ref{subsec:IV-E}. The methods compared below are ideal MFP, mismatched MFP, broadband MBIP, and the proposed TEDS method.

\subsection{\label{subsec:IV-B} Visualization of spectral-angular and temporal modes}

We first examine the modes extracted by TEDS at two representative source depths, $100\,\mathrm{m}$ and $200\,\mathrm{m}$, under three SNR conditions: noise-free, $-5$ and $-15\,\mathrm{dB}$. 

\begin{figure*}[t]
	\center
	\includegraphics[width=2\columnwidth]{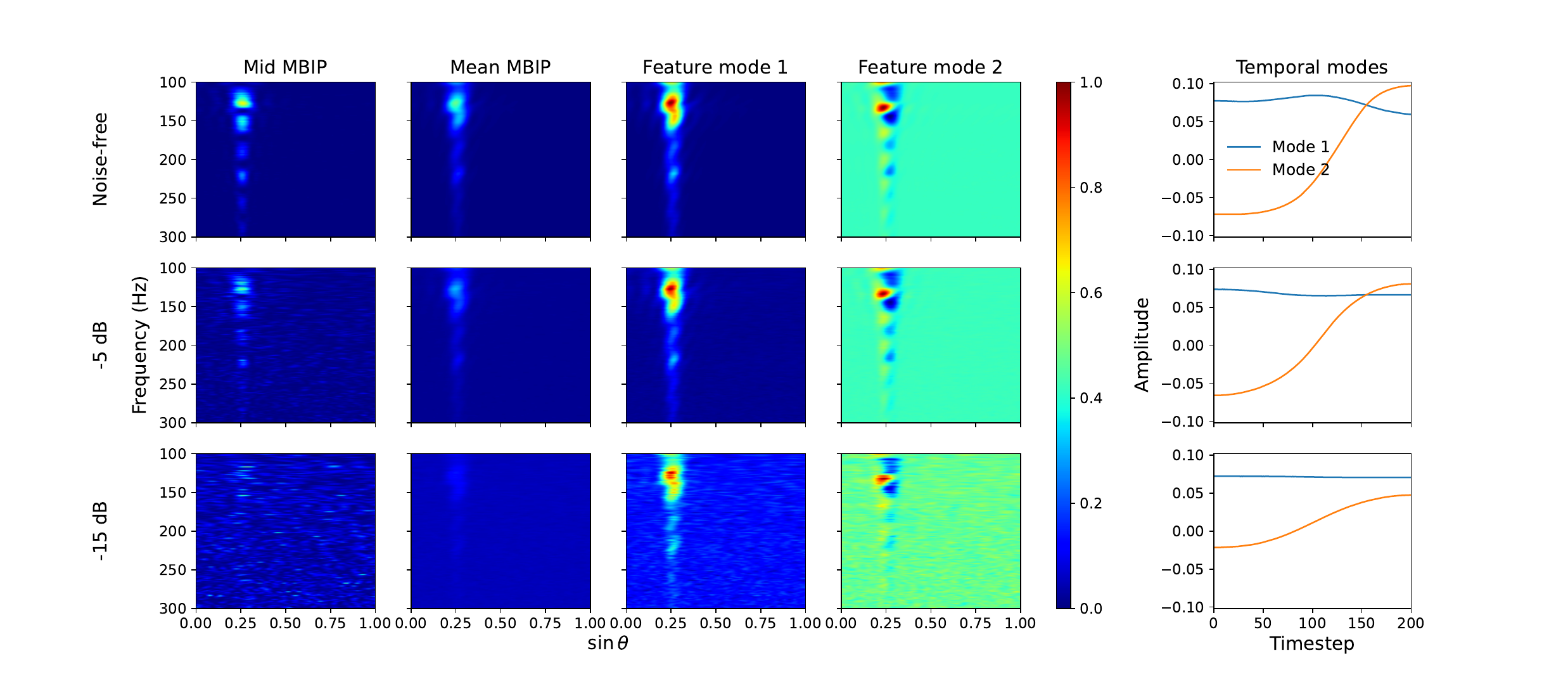}
	\caption{Broadband MBIP surfaces and TEDS-extracted modes for a $100\,\mathrm{m}$-deep source.
		Rows correspond to different SNR conditions (noise-free, $-5$ and $-15\,\mathrm{dB}$, from top to bottom). Columns show the MBIP beam-intensity surface in Eq.~\eqref{eq:beamforming} at the middle time instant (Mid MBIP), the time-averaged MBIP surface (Mean MBIP), the first spectral-angular mode, the second spectral-angular mode, and the corresponding temporal modes (from left to right).}
	\label{fig:modes_100m}
	\hrule
\end{figure*}

\begin{figure}[t]
	\centering
	\includegraphics[width=\linewidth]{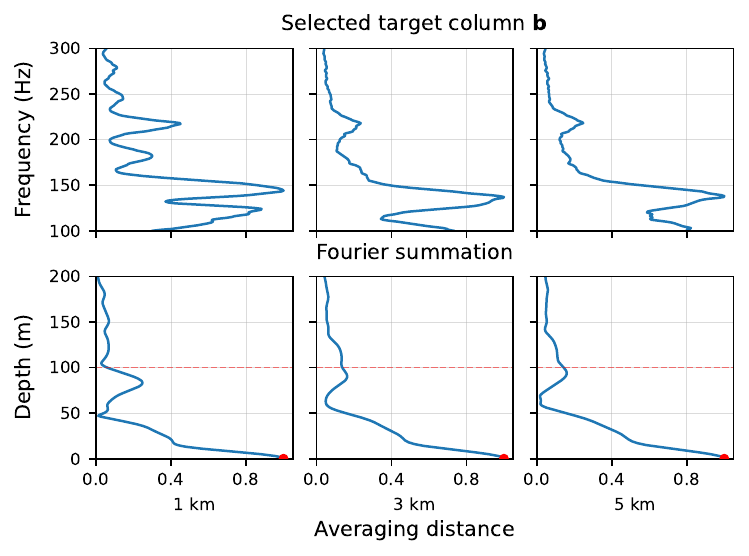}
	\caption{Temporal averaging baseline for a $100\hspace{0.166em}\mathrm{m}$-deep source. Top: selected target column $\mathbf{b}$ in \textbf{Algorithm~2}; bottom: Fourier summation in Eq.~\eqref{eq:fourier_sum}, with red dots marking the maxima.} 
	\label{fig:temporal_average}
\end{figure}

Fig.~\ref{fig:modes_100m} shows the results for a $100\,\mathrm{m}$-deep source. 
As the SNR decreases, the interference pattern in the instantaneous MBIP surface becomes less visible: it is clear with no noise, remains identifiable at $-5\,\mathrm{dB}$, and is difficult to discern at $-15\,\mathrm{dB}$. 
Although time averaging can suppress random fluctuations, it does not preserve the dynamic interference of a moving source because the target bearing shifts along the angular dimension. 
This is illustrated in Fig.~\ref{fig:temporal_average}, where Fourier summation is applied to noise-free time-averaged MBIP surfaces over different averaging distances. 
The selected target column $\mathbf{b}$ and the corresponding Fourier summation results show that direct temporal averaging blurs the depth-sensitive interference, leading to failed depth estimation.

TEDS avoids this limitation by extracting coherent modes from MBIP sequence. 
The first spectral-angular mode captures the average structure. 
It is physically similar to the Mean MBIP surface, with the main difference appearing in the relative energy scaling. 
Compared with the Mid MBIP surface, both the Mean MBIP surface and the first spectral-angular mode exhibit a more blurred interference structure, where neighboring bright regions tend to merge. 
This blurring is unfavorable for depth estimation because Fourier summation relies on a clearly separated interference periodicity.

In contrast, the second spectral-angular mode preserves the dominant depth-sensitive interference and remains interpretable even at low SNR. 
Its corresponding temporal coefficient varies sinusoidally, consistent with the temporal oscillation of the interference pattern as the source moves.\cite{mccargar2013depth} 
This indicates that the low-rank evolution model is able to suppress incoherent fluctuations while preserving the physically meaningful oscillatory structure used for depth estimation.

\begin{figure*}[t]
	\center
	\includegraphics[width=2\columnwidth]{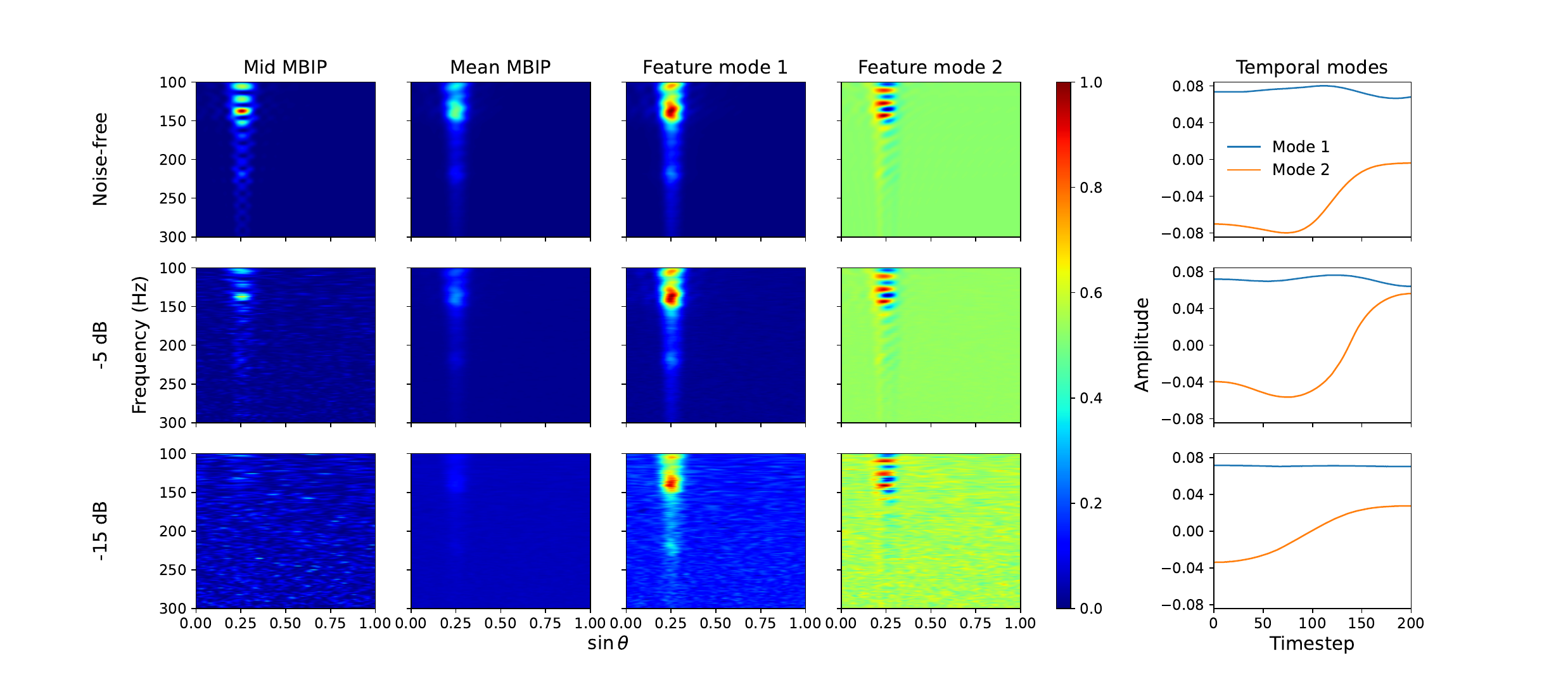}
	\caption{As Fig.~\ref{fig:modes_100m}, but at $200\,\mathrm{m}$ depth.}
	\label{fig:modes_200m}
	\hrule
\end{figure*}

The same analysis is repeated for a $200\,\mathrm{m}$-deep source in Fig.~\ref{fig:modes_200m}. 
Compared with the $100\,\mathrm{m}$ case, the interference contains more cycles over the same band, consistent with the depth dependence in Eq.~\eqref{eq:interference}. 
The qualitative modal structure remains similar. The first spectral-angular mode resembles the mean MBIP surface and represents the average energy over the observation interval, with a flat temporal mode. 
In contrast, the second spectral-angular mode retains clearer separated interference structure, with a sinusoidal temporal mode. 
This division of roles supports the interpretation that TEDS separates the mean component from the depth-sensitive interference and captures their temporal evolution.

\subsection{\label{subsec:IV-C} Depth localization results}

We next compare four representative methods---ideal MFP, mismatched MFP, broadband MBIP, and TEDS---under different SNR. Ten Monte Carlo trials are conducted for the $100\,\mathrm{m}$-deep source. Localization accuracy is quantified by the mean absolute error (MAE),\cite{willmott2005advantages}
\begin{align}
	\mathrm{MAE}=\frac{1}{10}\sum_{i=1}^{10}|\hat{z}_i-z_s|,
	\label{eq:mae}
\end{align}
where $\hat{z}_i$ is the estimated depth in the $i$-th trial and $z_s$ is the true source depth.

\begin{table}[t]
	\begin{center}
		\caption{MAEs of different methods under different SNR conditions (source depth = $100\,\mathrm{m}$).}
		\label{tab:mae_snr}
		\begin{ruledtabular}
			\begin{tabular}{c cccc}
				SNR (dB) & \textbf{MFP} & \textbf{Mismatched MFP} & \textbf{MBIP} & \textbf{TEDS} \\
				\hline 
				$\infty$ & 0.0  & 0.0  & 12.0 & 4.0  \\
				$-5$     & 0.0  & 5.0  & 17.5 & 3.6  \\
				$-10$    & 0.0  & 6.5  & 34.1 & 5.7  \\
				$-15$    & 0.25 & 21.0 & 44.2 & 6.3
			\end{tabular}
		\end{ruledtabular}
	\end{center}
\end{table}

Except for TEDS, which uses the full MBIP sequence over the track, the other methods are applied independently at each snapshot. 
Their final estimate is obtained by averaging the per-snapshot estimates. The results are summarized in Table~\ref{tab:mae_snr}. Ideal MFP provides the expected upper bound and remains nearly perfect. Once mismatch is introduced, its performance degrades with decreasing SNR, consistent with the well-known sensitivity of MFP to environmental and system mismatch.

Broadband MBIP is less sensitive to mismatch, but it relies on extracting periodic interference from individual observations. As the SNR decreases, the interference becomes increasingly difficult to recover directly, leading to substantial performance degradation. TEDS is more robust because it does not rely on a single snapshot. Instead, it exploits the coherent temporal evolution of MBIP features and suppresses incoherent fluctuations through the low-rank tensor-constrained model. As a result, TEDS consistently outperforms mismatched MFP and MBIP under all noisy conditions and remains stable even at $-15\,\mathrm{dB}$.

For TEDS, the residual error in the noise-free case is caused by mismatch between the dual-path model in Eq.~\eqref{eq:rap_pressure} and the full simulated acoustic field. 
The noise-free estimate is a biased reference produced by the simplified model. 
The results under noisy conditions then reflect deviations from this biased reference, rather than deviations from an exactly unbiased value. 
That the MAE at $-5\,\mathrm{dB}$ is slightly smaller than that in the noise-free case is understandable, considering the random noise realization and the limited number of Monte Carlo trials.
Weak noise perturbs the estimate away from the biased reference, but whether this perturbation moves the estimate closer to or farther from the ground-truth is trial-dependent. 
As the noise level increases, the extracted mode is less stable and the MAE increases, consistent with the performance trend.

\begin{figure}[!t]
	\center
	\includegraphics[width=1\columnwidth]{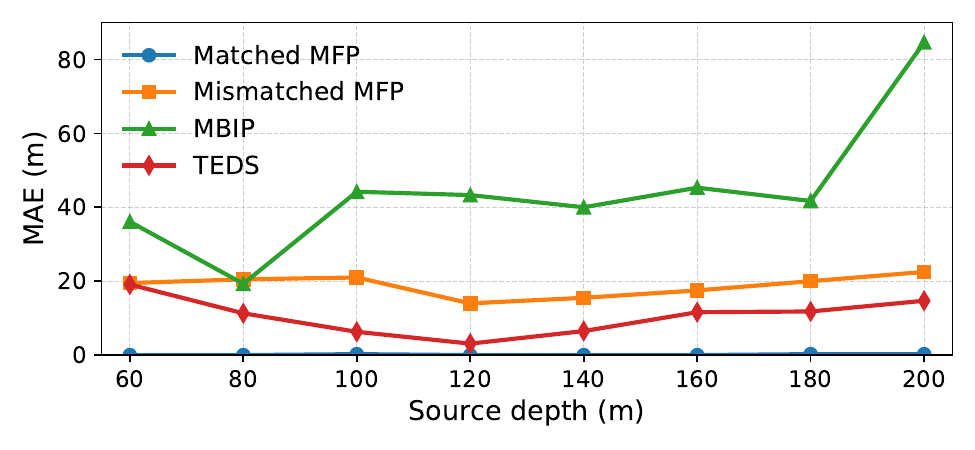}
	\caption{MAE of depth estimation at $-15\,\mathrm{dB}$ SNR versus source depth for ideal matched MFP, mismatched MFP, MBIP, and the proposed TEDS method.}
	\label{fig:compare4methods}
	\hrule
\end{figure}

To further examine robustness across source depth, Fig.~\ref{fig:compare4methods} shows the MAE at $-15\,\mathrm{dB}$ for source depths from $60$ to $200\,\mathrm{m}$. Ideal MFP remains nearly perfect. TEDS consistently outperforms mismatched MFP and MBIP over the full depth range. Its MAE first decreases and then increases with depth. This reflects two competing effects. As source depth increases from $60$ to $140\,\mathrm{m}$, the number of interference cycles within the same frequency band increases, which improves depth discriminability. At larger depths, however, the surface-reflected path traverses shallow layers with stronger sound-speed gradients, so the actual interference deviates more from the simplified model in Eq.~\eqref{eq:interference}. This limits the accuracy of RAP-based depth estimation, including TEDS.

Since TEDS is built on MBIP sequence extracted from a vertical array, array tilt directly affects the estimated source angle and distorts the spectral-angular interference. However, TEDS only uses sound speed in Fourier summation and does not explicitly rely on other environmental parameters. 
Previous MBIP studies suggest that the choice of sound speed does not need to be very strict, since the depth-estimation result is not sensitive to moderate sound-speed deviations.\cite{mccargar2013depth,zheng2020matched,zhou2022target} 
Therefore, array tilt is a dominant source of mismatch for TEDS, as discussed in Appendix~\ref{app:array_tilt}.

\subsection{\label{subsec:IV-D} Computational cost}

We compare the computational cost of MFP, broadband MBIP, and TEDS using the $100\,\mathrm{m}$-deep source.
The runtimes are measured on a computer with an Intel i7 CPU at $2.20\,\mathrm{GHz}$ and $32\,\mathrm{GB}$ memory.

For MFP, the main cost is replica generation.
Using Kraken in MATLAB, generating replicas for $40$ frequency samples over the range-depth search grid takes about $3\,\mathrm{h}$.
Once the replicas are precomputed, loading them and performing grid matching takes only about $3\,\mathrm{s}$.
Therefore, practical MFP depth estimation necessarily relies on precomputed replicas.
Since these replicas are tied to assumed environmental parameters, environmental mismatch is unavoidable.
As shown in Sec.~\ref{subsec:IV-C}, TEDS achieves better depth-estimation accuracy than mismatched MFP.

Broadband MBIP avoids replica generation by comparing observed interference with predicted one, see Eq.~\eqref{eq:mbip}.
Its computation consists of MBIP construction followed by matching operation, which takes about $15\,\mathrm{s}$.
This makes broadband MBIP computationally efficient, but its single-snapshot processing is more sensitive to noise.

TEDS uses the MBIP sequence over the source track.
Let $N_f$ be the number of frequency samples, $N_\theta$ the number of angular samples, $T$ the number of time frames, and $N=N_fN_\theta$ the feature dimension.
In this experiment, $N_f=200$, $N_\theta=200$ and $T=200$.
In one outer iteration, the dominant operations are products between the MBIP sequence and the low-rank factors, giving a leading cost of
\begin{equation}
	O\left((1+L_V+L_X)NRT\right),
\end{equation}
With small rank $R$, the computation scales mainly linearly with the feature dimension $N$ and the sequence length $T$.

In our accelerated implementation, the main matrix products and tensor contractions are computed using PyTorch operations such as \texttt{torch.matmul} and \texttt{torch.einsum},\cite{paszke2019pytorch}
The resulting runtime is $25\,\mathrm{s}$.
The main latency comes from data accumulation rather than computation: at a source speed of $5\,\mathrm{m/s}$, a $2\,\mathrm{km}$ track requires $400\,\mathrm{s}$ of observations.
In continuous tracking, the latency can be reduced by updating TEDS with newly acquired observations and historical data.

\subsection{\label{subsec:IV-E} Impact of tracking duration}

Tracking duration has a direct influence on the quality of mode recovery. 
A longer track provides more temporal information, but it introduces stronger bearing migration. 
As a result, the depth-sensitive interference pattern and the target-motion pattern are distributed across different modes.

\begin{figure}[t]
	\center
	\includegraphics[width=1\columnwidth]{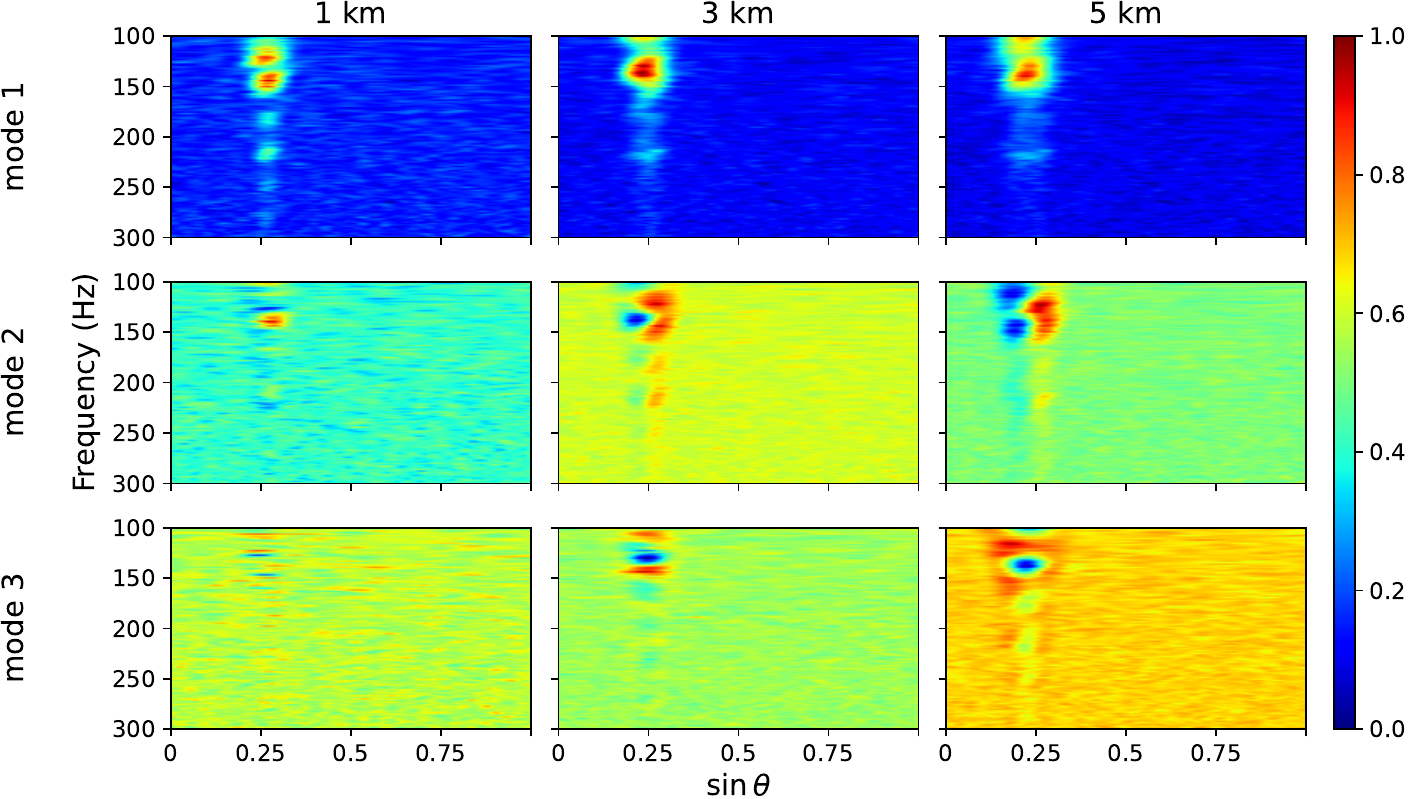}
	\caption{Spectral-angular modes extracted by TEDS for 1-$5\,\mathrm{km}$ tracking durations at $-15\,\mathrm{dB}$ SNR for a $100\,\mathrm{m}$-deep source. Each column corresponds to one tracking duration, and each row shows spectral-angular modes~1--3.}
	\label{fig:subsec_D}
	\hrule
\end{figure}

All experiments in this subsection are conducted at $-15\,\mathrm{dB}$, with a source depth of $100\,\mathrm{m}$. 
Table~\ref{tab:tracking_duration_mae} reports the MAE from ten Monte Carlo trials for tracking durations from $1$ to $5\,\mathrm{km}$, and the mode evolution is shown in Fig.~\ref{fig:subsec_D}.
Depth estimation is performed by applying Fourier summation to spectral-angular modes~2 and~3.

\begin{table*}[t]
	\begin{center}
		\caption{MAEs under different tracking durations at $-15\,\mathrm{dB}$ SNR using modes 2 and 3 (source depth = $100\,\mathrm{m}$).}
		\label{tab:tracking_duration_mae}
		\begin{ruledtabular}
			\begin{tabular}{c ccccccccc}
				Tracking duration ($\mathrm{km}$) & 1 & 1.5 & 2 & 2.5 & 3 & 3.5 & 4 & 4.5 & 5 \\
				\hline
				MAE of Mode 2 ($\mathrm{m}$) & 14.0 & 15.1 & 6.3 & 4.4 & 4.4 & 4.0 & 2.8 & 25.5 & 79.5 \\
				MAE of Mode 3 ($\mathrm{m}$) & 44.8 & 17.8 & 11.8 & 11.1 & 8.7 & 8.4 & 8.5 & 8.2 & 7.1
			\end{tabular}
		\end{ruledtabular}
	\end{center}
\end{table*}

For mode~2, the MAE first decreases and then increases with tracking duration. 
When the track is too short, temporal accumulation is limited and the extracted mode is less stable. 
For durations between $2$ and $4\,\mathrm{km}$, the MAE remains below $10\,\mathrm{m}$, indicating that mode~2 contains a clear depth-sensitive interference structure. 
However, as the track becomes longer, the target-bearing migration becomes more pronounced and competes with the depth-related interference in the lower modes, see mode~2 at $5\,\mathrm{km}$ in Fig.~\ref{fig:subsec_D}, where the periodic interference is split into two bearing columns.
This explains the sharp degradation of mode~2 beyond $4\,\mathrm{km}$.

In contrast, mode~3 becomes increasingly useful for longer tracking durations. 
Although it is less reliable for short tracks, its MAE decreases as more temporal information becomes available and remains below $10\,\mathrm{m}$ for tracking durations longer than $3\,\mathrm{km}$. 
This behavior indicates that, when the source motion induces a noticeable bearing shift, mode~2 partly captures the bearing-migration component, while the depth-sensitive interference is more clearly isolated in mode~3. Therefore, the use of $R=3$ provides an additional degree of freedom to separate the mean component, the motion-related component, and the depth-sensitive interference.

\begin{figure}[!t]
	\center
	\includegraphics[width=1\columnwidth]{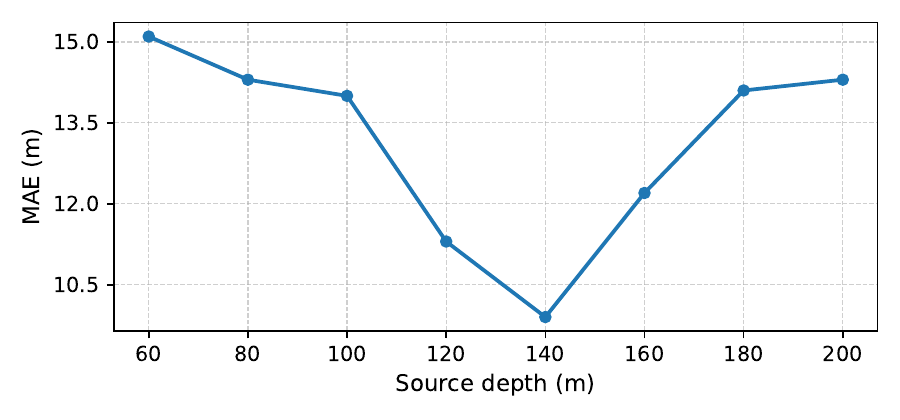}
	\caption{MAE of depth estimation versus source depth using TEDS with a tracking duration of $1\,\mathrm{km}$ under $-15\,\mathrm{dB}$ SNR.}
	\label{fig:track_1km}
	\hrule
\end{figure}

To assess performance under a more stringent rapid-localization requirement, depth estimation is performed using a fixed tracking duration of $1\,\mathrm{km}$ for source depths from $60$ to $200\,\mathrm{m}$, as shown in Fig.~\ref{fig:track_1km}. 
As the source depth increases to $140\,\mathrm{m}$, the MAE decreases, reflecting improved depth discriminability due to the larger number of interference cycles.
As the depth further increases to $200\,\mathrm{m}$, the MAE begins to rise. 
This again indicates a limitation of RAP-based depth estimation: for deeper sources, the simplified depth-dependent interference model in Eq.~\eqref{eq:rap_pressure}--\eqref{eq:interference} becomes less consistent with the actual propagation field.
Even so, the MAE remains below $15\,\mathrm{m}$ over the entire depth range, indicating that TEDS remains robust even with limited temporal information.

\section{\label{sec:5} Conclusions and Future Directions}

This paper presented a tensor evolution-based depth estimation (TEDS) method for shallow-source depth estimation using a deep-ocean near-bottom vertical line array. TEDS models broadband MBIP feature evolution by a time-varying autoregressive operator with a low-rank Tucker structure, so that the dominant depth-dependent interference can be extracted from a small number of spectral-angular modes and temporal modes.

The results show that TEDS improves robustness for low-SNR and consistently outperforms mismatched MFP and broadband MBIP. This suggests that treating RAP-based depth estimation as a dynamic feature-evolution problem, rather than a snapshot-based one, is beneficial in noise.

The framework is inspired by the DMD viewpoint, but extends it to a time-varying setting suitable for MBIP sequence. Its tensor modeling strategy is also distinct from most previous ocean-acoustic applications, in that the tensor structure is imposed on the evolution process rather than on a static field quantity.

The linear time-varying evolution model used in TEDS is a local approximation in the MBIP feature domain. The approximation may degrade for long tracks with strong bearing migration, low SNR where coherent spectral-angular modes cannot be extracted, or deeper sources for which the actual propagation delay deviates from the simplified RAP model.

Future work will therefore consider nonlinear or physics-informed extensions to improve the flexibility of the model. Possible directions include propagation-constrained operator learning, adaptive mode-selection strategies for strong source-motion cases, and the incorporation of array perturbations into the evolution model to better handle system mismatch.

\section{Author Declarations}
The authors have no conflict of interest to disclose.

\section{Data Availability}
This paper is based on simulated data. The Python implementation of TEDS used in this study is publicly available at \url{https://github.com/OceanSTARLab/TEDS}.

\appendix

\section{Sensitivity analysis}
\label{app:sensitivity}

The tensor rank $R$ and the TV weight $\lambda$ are the two main user-specified parameters in TEDS. This appendix summarizes their effects on depth estimation and gives the parameter choices used in this work.

The rank $R$ controls the dimension of the latent evolution subspace. A small rank is preferred if it can provide accurate depth estimation, since a larger rank increases the computational cost. As discussed in Sec.~\ref{subsec:IV-B}, Spectral-angular mode~1 mainly represents the average energy over the observation interval and has a blurred interference structure. Thus, it cannot provide a reliable depth estimate. Sec.~\ref{subsec:IV-E} shows that, for the short-track case, mode~2 gives the best performance. For longer tracks, however, mode~2 partly captures the bearing-evolution component, while the depth-sensitive interference pattern appears more clearly in mode~3.
Therefore, we choose $R=3$ rather than the minimum choice $R=2$. This provides one additional mode to separate motion-induced feature migration from depth-sensitive interference, while keeping the model compact.

We next examine the influence of the TV weight $\lambda$ with $R=3$. The parameter $\lambda$ controls the smoothness of the temporal modes. When $\lambda=0$, the temporal modes are constrained only by the low-rank evolution model and are more sensitive to noise.
As shown in Table~\ref{tab:lambda_sensitivity}, a moderate TV penalty improves the depth-estimation accuracy. Therefore, $\lambda=0.1$ is selected as a stable value that suppresses temporal fluctuations without over-smoothing the coherent evolution.

\begin{table}[t]
	\begin{center}
		\caption{Sensitivity of TEDS to the TV weight $\lambda$. The entries are MAEs $(\mathrm{m})$ obtained using mode~2.}
		\label{tab:lambda_sensitivity}
		\begin{ruledtabular}
			\begin{tabular}{c ccccc}
				SNR & $\lambda=0$ & $0.01$ & $0.1$ & $1$ & $10$ \\
				\hline
				$-5\hspace{0.166em}\mathrm{dB}$ & 12.0 & 5.2 & 3.6 & 3.7 & 4.9 \\
				$-15\hspace{0.166em}\mathrm{dB}$ & 15.6 & 5.8 & 6.3 & 6.4 & 6.7
			\end{tabular}
		\end{ruledtabular}
	\end{center}
\end{table}
	
\section{Initialization Strategy}
\label{app:algorithm_parameters}

This appendix gives the initialization strategy used in the reported experiments. Before the alternating updates, the snapshot matrix is formed from the MBIP feature sequence as $\mathrm{Y}=\mathrm{Y}_1$ in Eq.~~\eqref{eq:dmd_snapshots}.
A rank-$R$ truncated singular value decomposition is then applied,
\begin{align}
	\mathbf{Y}
	\approx
	\mathbf{U}_R\boldsymbol{\Sigma}_R\mathbf{Q}_R^T,
\end{align}
where $\mathbf{U}_R\in\mathbb{R}^{N\times R}$, $\boldsymbol{\Sigma}_R\in\mathbb{R}^{R\times R}$, and $\mathbf{Q}_R\in\mathbb{R}^{(T-1)\times R}$. The two feature-domain factor matrices are initialized by the leading left singular vectors,
\begin{align}
	\label{ini_w_v}
	\mathbf{W}^{(0)}=\mathbf{U}_R,\qquad
	\mathbf{V}^{(0)}=\mathbf{U}_R .
\end{align}
The temporal factor is initialized by the scaled right singular vectors,
\begin{align}
	\label{ini_x}
	\mathbf{X}^{(0)}=\mathbf{Q}_R\boldsymbol{\Sigma}_R .
\end{align}
Thus, the $t$-th row of $\mathbf{X}^{(0)}$ provides the initial temporal coefficient $\mathbf{x}_t^{(0)T}$. Given $\mathbf{W}^{(0)}$, $\mathbf{V}^{(0)}$, and $\mathbf{X}^{(0)}$, the unfolded core tensor is initialized by least squares, following Eq.~\eqref{eq:updateG}.

\section{Robustness to array tilt}
\label{app:array_tilt}

We first analyze how array tilt affects the MBIP feature used by TEDS. 
For the vertical array considered in Sec.~\ref{subsec:II-A}, the beamforming in Eq.~\eqref{eq:beamforming} uses the phase factor 
$e^{ik(jd-\bar{z})\sin\theta}$, which assumes that the array elements are aligned vertically. When the array is tilted by an angle $\alpha$ in radians, the effective angular term becomes
\begin{equation}
	\sin(\theta_s+\alpha)
	\approx
	\sin\theta_s+\alpha\cos\theta_s ,
	\label{eq:tilt_angle_shift}
\end{equation}
The induced angular bias is therefore
\begin{equation}
	\Delta(\sin\theta_s)
	\approx
	\alpha\cos\theta_s .
	\label{eq:tilt_angle_bias}
\end{equation}
Since the RAP-induced depth information in Eq.~\eqref{eq:interference} depends on the phase term $2kz_s\sin\theta_s$, this angular bias perturbs the interference periodicity used by Fourier summation and shifts the depth estimate.

We evaluate this effect using a $100\hspace{0.166em}\mathrm{m}$-deep source.  
Three SNR conditions are tested: noise-free, $-5\hspace{0.166em}\mathrm{dB}$, and $-15\hspace{0.166em}\mathrm{dB}$. For comparison, MFP is tested under the same array-tilt without additional mismatch, using 40 uniformly selected frequencies and searches over the depth-range grid.

\begin{figure}[t]
	\center
	\includegraphics[width=1\columnwidth]{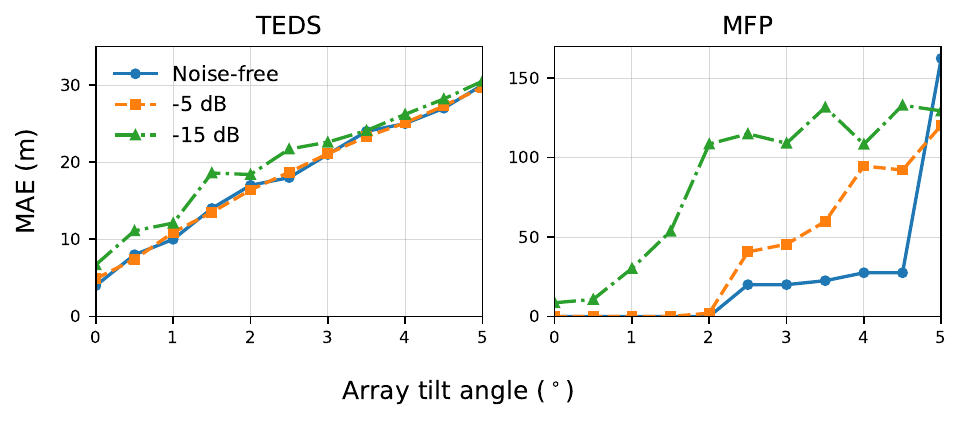}
	\caption{Depth-estimation MAE of TEDS and MFP under array-tilt mismatch for a $100\hspace{0.166em}\mathrm{m}$-deep source.}
	\label{fig:array_tilt_compare}
	\hrule
\end{figure}

Fig.~\ref{fig:array_tilt_compare} shows that the MAE of TEDS increases almost linearly with the tilt, consistent with Eq.~\eqref{eq:tilt_angle_bias}. For the same tilt, the errors under different SNR conditions are close to each other. This indicates that the dominant error source in this test is the array tilt, while the extracted depth-sensitive mode is stable. The nonzero error at $0^\circ$ comes from the mismatch between the simplified model and the actual propagation field.

MFP is accurate for small tilt and moderate SNR. 
However, its error increases sharply as the tilt or noise increases. 
This MFP test is idealized, since only tilt is considered. MFP is also affected by sound-speed, seabed, and other replica-model errors. 
In contrast, TEDS only uses sound speed in Fourier summation step, so its sensitivity to environmental parameters is weaker.

\bibliography{tedsbib_new}

\end{document}